\documentclass[11pt,a4paper]{article}
\usepackage{jheppub}
\usepackage{amsmath}
\usepackage{graphicx}
\usepackage{color}
\usepackage{bm}
\usepackage{epstopdf}
\begin{document}

\title{Static $\bar{Q}Q$ pair free energy and screening masses from correlators of Polyakov loops: continuum extrapolated lattice results at the QCD physical point}
\author[1]{Szabolcs Bors\'anyi,}
\author[1,2,3]{Zolt\'an Fodor,}
\author[3,4]{S\'andor D. Katz,}
\author[3,4]{Attila P\'asztor,}
\author[1,2]{K\'alm\'an K. Szab\'o,}
\author[3]{Csaba T\"or\"ok}

\affiliation[1]{University of Wuppertal, Department of Physics, Wuppertal D-42097, Germany}
\affiliation[2]{J\"ulich Supercomputing Center, J\"ulich D-52425, Germany}
\affiliation[3]{E\"otv\"os University, Budapest H-1117, Hungary}
\affiliation[4]{MTA-ELTE Lend\"ulet Lattice Gauge Theory Research Group}

\abstract{
We study the correlators of Polyakov loops, and the corresponding gauge invariant free energy of a
static quark-antiquark pair in 2+1 flavor QCD at finite temperature. Our simulations were 
carried out on $N_t=6,8,10,12,16$ lattices using Symanzik improved gauge action and a stout 
improved staggered action with physical quark masses. The free energies calculated from the Polyakov loop 
correlators are extrapolated to the continuum limit. For the free energies we use a two step renormalization 
procedure that only uses data at finite temperature. We also measure correlators with definite Euclidean 
time reversal and charge conjugation symmetry to extract two different screening masses, 
one in the magnetic, and one in the electric sector, to distinguish two different correlation
lengths in the full Polyakov loop correlator.
}

\maketitle

\section{Introduction}

At high temperatures strongly interacting matter undergoes a
transition where colorless hadrons turn into a phase dominated by
colored quarks and gluons, the quark gluon plasma (QGP). Recently, lattice simulations 
have shown that this transition is a crossover~\cite{Aoki:2006we} and its
characteristic temperature has also been 
determined~\cite{Cheng:2006qk,Aoki:2006br,Aoki:2009sc,Borsanyi:2010bp,Bazavov:2011nk}.
Deconfinement properties of the transition can be studied by
infinitely heavy, static test charges. At zero temperature a heavy quark and antiquark pair forms a 
bound state (quarkonium state), but above the deconfinement temperature, 
color screening and collisions with the medium would reduce the binding between the quark 
and the antiquark, eventually causing a dissociation. 
As proposed by Ref. ~\cite{Matsui:1986dk}, 
the behavior of quarkonia can signal deconfinement and QGP production 
in heavy ion experiments. Moreover, the different melting temperatures of the different states
can be used as a thermometer, analogously to the spectral analysis of stellar media in astrophysics, 
where the absence and presence of the different spectral lines is used to determine the temperature.\\

In medium quarkonium properties are characterized by the corresponding spectral functions, 
studied in several works. However, extracting spectral functions from Euclidean meson 
correlators (i.e. the analytic continuation of the correlator to real time) is a difficult, 
ill-posed problem. Nevertheless, lattice studies of charmonium spectral functions 
using the Maximum Entropy Method have been carried out on numerous 
occasions \cite{Jakovac:2006sf, Umeda:2002vr, Asakawa:2003re, Iida:2006mv, Ohno:2011zc, Ding:2012sp, Aarts:2007pk, Kelly:2013cpa, Borsanyi:2014vka, Aarts:2012ka, Aarts:2013kaa, Kim:2013seh}. 
A recent, detailed study of charmonium spectral functions in 
quenched QCD can be found in \cite{Ding:2012sp}. Results regarding spectral functions 
with 2 flavours of dynamical quarks can be found in Refs. \cite{Aarts:2007pk, Kelly:2013cpa}. 
A recent study of charmonium spectral functions in 2+1 flavour QCD is \cite{Borsanyi:2014vka, Borsanyi:2014pta}. Bottomonium
spectral functions have also been studied with the help of 
NRQCD \cite{Aarts:2012ka, Aarts:2013kaa,Kim:2013seh}. \\ 

Since the direct determination of the spectral function is difficult, one can study 
in-medium properties of quarkonium using approximate potential models.
There are numerous proposals in the literature for lattice observables which can provide
input to these models. The so-called singlet and octet potentials have been 
proposed \cite{Nadkarni:1986as, Kaczmarek:2002mc, Kaczmarek:2004gv, Kaczmarek:2005ui, Mocsy:2007yj}, 
and studied on the lattice, but these are not gauge invariant, therefore 
extracting physical information from them is not straightforward. There was also a suggestion about 
using the analytic continuation of the Wilson-loop \cite{Laine:2006ns, Rothkopf:2011db}, that, however, 
has similar problems as the direct reconstruction of the spectral functions. Here, we calculate the gauge 
invariant static quark-antiquark pair free energy, a non-perturbatively well defined quantity, 
that carries information on the deconfinement properties of the QGP.\\

In the present paper we determine the free energy of a static quark-antiquark pair
as a function of their distance at various temperatures. We accomplish it by measuring the 
Polyakov loop correlator~\cite{McLerran:1981pb}, which gives the 
gauge invariant $\bar{Q}Q$ free energy \footnote{More precisely, the excess free energy that 
we get when inserting two static test charges in the medium.} as:

\begin{equation} \label{def_of_F}
F_{\bar{Q}Q}(r) = -T \ln C(r,T) = - T \ln \left\langle \sum_{\textbf{x}} \mathrm{Tr} L(\textbf{x}) \mathrm{Tr} L^+(\textbf{x}+\textbf{r}) \right\rangle \rm{.}
\end{equation}
In the above formula, $\textbf{x}$ runs over
all the lattice spatial sites, and the Polyakov loop, $L(\textbf{x})$,
is defined as the product of temporal link variables $U_4(\textbf{x},x_4) \in SU(3)$\footnote{In the literature, a factor of
$\frac{1}{N_c}$ is often included in the definition. Including this factor leads to a term in the static quark free
energy that is linear in temperature.}:
\begin{equation}
L(\textbf{x})=\prod_{x_4=0}^{N_t-1} U_4 (\textbf{x},x_4) \rm{,}
\end{equation}
or in the continuum formulation, as a path ordered exponential of the integral of the gauge fields:
\begin{equation}
L_{\rm{cont}}(\textbf{x}) = \mathcal{P} e^{i g \int_0^{1/T} A_4(\tau, \textbf{x}) d \tau}  \rm{.}
\end{equation}
The leading order term to the correlator of Polyakov loops is
a two gluon exchange diagram. It was first calculated at leading order
in the dimensionally reduced effective theory (EQCD$_3$)~\cite{Nadkarni:1986cz}. 
Due to the two gluon exchange, the $r$ dependence in leading order 
is ${\mathrm{exp}}(-2m_D r)/r^2$, where $m_D$ is the Debye screening mass. This suggests 
that in the $r \to 0$ limit, where perturbation
theory is applicable, the correlator should behave as $1/r^2$. However, this is not the $r \to \infty$ asymptotic 
behavior, which we need to fit the correlation length on the lattice. The reason is simple:
even in the weak coupling limit, at distances larger than $(g^2 T)^{-1}$ the physics of magnetic screening
becomes dominant. From the then applicable 3D effective pure Yang-Mills theory, Ref.~\cite{Braaten:1994qx} argued 
that at high temperature, the large distance behavior is ${\mathrm{exp}}(-m_M r)/r$, where $m_M$ is the magnetic 
screening mass. This was confirmed by 2 flavour lattice simulations (using a somewhat heavy pion) in~\cite{Maezawa:2010vj}.\\

A related problem is that for the gluon self-energy, perturbation
theory breaks down at the $\mathcal{O}(g^2 T)$ order because of infrared divergences. This term 
contains contributions from magnetic gluons. Therefore, the perturbative definition
of the screening mass, as a pole in the gluon propagator, is of limited use, since pertubation 
theory breaks down (\cite{Arnold:1995bh}). It is better to define the screening masses as inverse correlation lengths 
in appropriate Euclidean correlators. In order to investigate the effect of electric and magnetic gluons separately, 
one can use the symmetry of Euclidean time reflection~\cite{Arnold:1995bh}, 
that we will call ${\cal R}$. The crucial property of magnetic versus
electric gluon fields $A_4$ and $A_i$ is that under this symmetry, one is intrinsically 
odd, while the other is even:
\begin{equation}
 A_4(\tau,\textbf{x}) \xrightarrow{{\cal R}} - A_4(-\tau,\textbf{x})\rm{,} \quad A_i(\tau,\textbf{x}) \xrightarrow{{\cal R}} A_i(-\tau,\textbf{x})
\end{equation}
Under this symmetry the Polyakov loop transforms as $L \xrightarrow{{\cal R}} L^{\dagger}$. One can easily define correlators that are even or odd under 
this symmetry, and thus receive contributions only from the magnetic or electric sector, 
respectively~\cite{Arnold:1995bh, Maezawa:2010vj}: \\
\begin{eqnarray}
 L_M \equiv (L + L^\dagger)/2 \\
 L_E \equiv (L - L^\dagger)/2 \rm{.}
\end{eqnarray}
We can further decompose the Polyakov loop into ${\cal C}$ even and odd states, using 
$A_4 \xrightarrow{{\cal C}} A_4^*$ and $L \xrightarrow{{\cal C}} L^*$ as:
\begin{eqnarray}
L_{M\pm} = (L_M \pm L^*_M)/2 \\
L_{E\pm} = (L_E \pm L^*_E)/2 \rm{.}
\end{eqnarray}
Next, we note that $\operatorname{Tr} {L_{E+}}=0=\operatorname{Tr} {L_{M-}}$, so the decomposition of the Polyakov loop correlator to 
definite ${\cal R}$ and ${\cal C}$ symmetric operators contains two parts\footnote{Note that the Polyakov loop correlator does 
not overlap with the ${\cal R}({\cal C})=+(-)$ and ${\cal R}({\cal C})=-(+)$ sectors. To access these sectors, other
operators are needed.}. We define the magnetic correlation function as:
\begin{equation}
\label{eq:Cmag}
C_{M+}(r,T) \equiv \left\langle \sum_{\bf x} \mathrm{Tr} L_{M+}({\bf x}) \mathrm{Tr} L_{M+}({\bf x}+{\bf r}) \right\rangle - \left| \left\langle \sum_{\bf x} \mathrm{Tr} L({\bf x}) \right\rangle \right|^2 \rm{,}
\end{equation} 
\\
and the electric correlator as\footnote{Here our definition differs from that used in~\cite{Maezawa:2010vj} in a sign.}:
\\
\begin{equation}
\label{eq:Cel}
C_{E-}(r,T) \equiv - \left\langle \sum_{\bf x} \mathrm{Tr} L_{E-}({\bf x}) \mathrm{Tr} L_{E-}({\bf x}+{\bf r}) \right\rangle \rm{.}
\end{equation}

Then, from the exponential decay of these correlators, we can define the magnetic 
and electric screening masses. Note that with our definition $\mathrm{Tr}L_{M+} = \operatorname{Re} \mathrm{Tr}L$  and 
$\mathrm{Tr}L_{E-} = i \operatorname{Im} \mathrm{Tr}L$ , and:
\begin{equation}
C(r,T) - C(r \to \infty,T) = C_{M+}(r,T) + C_{E-}(r,T) \rm{,}
\end{equation}
from which it trivially follows that if the magnetic mass screening mass is lower than the electric mass, we will 
have $C(r,T) - C(r \to \infty,T)$ asymptotic to $C_{M+}(r,T)$ as $r \to \infty$, or equivalently, the highest 
correlation length in $C$ equal to that of $C_{M+}$. \\

As for the asymptotic form of these correlators, similar arguments apply as with the full Polyakov 
loop correlator. In the high temperature limit the asymptotic behavior will be dominated by a glueball mass in the 3D
effective Yang-Mills theory \cite{Arnold:1995bh, Braaten:1994qx}, but because of the symmetry properties, the quantum numbers carried by the glueballs will
be different. We will therefore fit the ansatz:
\begin{eqnarray}
C_{M+}(r,T) \xrightarrow{r \to \infty} K_{M}(T) \frac{\mathrm{e}^{-m_M(T)r}}{r} \rm{,}\\
C_{E-}(r,T) \xrightarrow{r \to \infty} K_{E}(T) \frac{\mathrm{e}^{-m_E(T)r}}{r} \rm{,}
\end{eqnarray}
to extract screening masses, noting that the ansatz in principle is only motivated at high temperatures, 
where the effective field theory applies. Nevertheless we find that even close to $T_c$ 
the ansatz describes the large $r$ tails of our lattice data well.

\section{Simulation details}
The simulations were performed by using the tree level Symanzik improved gauge, 
and stout-improved staggered fermion action, that was used in~\cite{Aoki:2005vt}.
We worked with physical quark masses, and fixed them by reproducing the 
physical ratios $m_{\pi}/f_K$ and $m_K/f_K$~\cite{Aoki:2005vt}. \\

\begin{figure}[h!]
\begin{center}
  \includegraphics[width=0.7\textwidth]{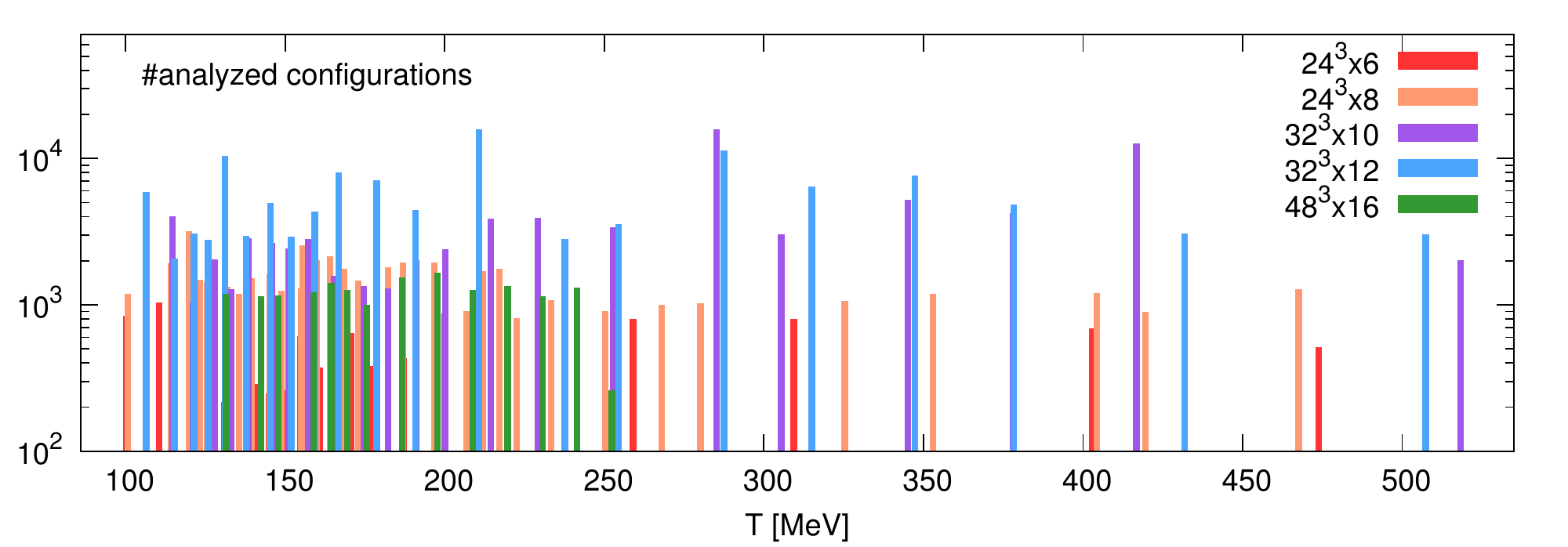}
\end{center}
\vspace{-0.5cm}
\caption{Number of the analyzed lattice configurations.}
\end{figure}

Compared to our previous investigations of Polyakov loop correlators, reported 
in the conference proceedings ~\cite{Fodor:2007mi}, here we used finer lattices, 
namely we carried out simulations on $N_t=12$ and $16$ lattices as well as 
on $N_t=6, 8, 10$ lattices. Our results were obtained in the temperature 
range 150 MeV $\leq$ T $\leq$ 450 MeV. We use the same configurations as in 
Ref. \cite{Borsanyi:2010bp} and \cite{Borsanyi:2013bia}.
Figure 1 summarizes our statistics.

\section{The gauge invariant free energy}

\subsection{Renormalization procedure and continuum extrapolation}

After measuring the Polyakov loop correlator $C(r,T)$ at $T=1/(N_ta)$ temperature, we 
computed the unrenormalized free energy according to $F_{\bar{Q}Q}=-T \ln C(r,T)$.
 The $a(\beta)$ function was taken from the 
line of constant physics, along which we kept the ratios of the physical 
values of $m_\pi$, $f_K$ and $m_K$ fixed at zero temperature. Detailed 
description of the determination of the line of constant physics can be 
found in Ref.~\cite{Borsanyi:2013bia}. \\

Approaching the continuum limit, the value of the unrenormalized free
energy diverges. In order to eliminate the divergent part of the free
energy renormalization is needed. 
There are various proposals in the literature for this renormalization procedure.
Earlier works~\cite{Kaczmarek:2002mc, Kaczmarek:2004gv, Kaczmarek:2005ui} matched 
the short distance behavior to the $T=0$ static potential, but this is ambiguous.
A more precise definition is to require that the $T=0$ potential
vanishes at some distance~\cite{Aoki:2006br, Fodor:2007mi}. 
This would require a precise determination of the potential at $T=0$. Here, 
though, we use a renormalization procedure based entirely on our $T>0$ data, similarly to Ref.  ~\cite{WilsonThermo}. 
The data contains a temperature independent divergent part from the ground state energy. The difference between the 
value of free energies at different temperatures is free of divergences.
Accordingly, we define the renormalized free energy as:

\begin{equation} \label{eq:renorm}
F_{\bar{Q}Q}^{ren} (r,\beta,T;T_0) = F_{\bar{Q}Q} (r,\beta,T) - F_{\bar{Q}Q}(r \rightarrow \infty, \beta, T_0) \rm{,}
\end{equation} 
\\
with a fixed $T_0$. This renormalization prescription corresponds to the choice that
the free energy at large distances goes to zero at $T_0$, and is implemented in two steps. In the first step we have:
\begin{equation}
\label{eq:Ftilde}
\tilde{F}_{\bar{Q}Q}(r,\beta,T) = F_{\bar{Q}Q} (r,\beta,T) - F_{\bar{Q}Q} (r \to \infty,\beta,T)  = F_{\bar{Q}Q} (r,\beta,T) - 2 F_{Q} (\beta,T)  \rm{,}
\end{equation}
where the one quark free energy $F_Q$ satisfies:
\begin{equation}
\label{eq:F1Q}
2 F_{Q} (\beta, T) = F_{\bar{Q}Q} (r \to \infty,\beta,T) = -T \log \left| \left< \operatorname{Tr} L  \right>\right|^2 \rm{.}
\end{equation}
Note, that this first step of the renormalization procedure is completely straightfoward to implement, at
each simulation point in $N_t$ and $\beta$ we just subtract the asymptotic value of the correlator.
This gives us a UV finite quantity $\tilde{F}_{\bar{Q}Q}$, however we don't 
call this the renormalized free energy, since compared to equation ($\ref{eq:renorm}$) it contains less information. Namely 
at all temperatures $\tilde{F}_{\bar{Q}Q}(r=\infty,T) = 0$. The correlation
length is the same as with definition ($\ref{eq:renorm}$), but the information of the asymptotic value (that is the single heavy quark
free energy) is lost. That information however is retained in the second step:
\begin{equation}
F_{\bar{Q}Q}^{ren}(r,\beta,T;T_0) = \tilde{F}_{\bar{Q}Q}(r,\beta,T) + 2 F^{ren}_Q (\beta, T; T_0) \rm{,}
\end{equation}
where the renormalized one heavy quark free energy is:
\begin{equation}
\label{eq:renorm1Q}
F^{ren}_Q(\beta,T; T_0) = F_{Q}(\beta,T)-F_{Q}(\beta,T_0) \rm{.}
\end{equation}
Doing the renormalization in two steps like this has a technical reason that will be explained shortly. \\

\begin{figure}[t!]
\begin{center}
\includegraphics[width=0.7\textwidth, angle=0]{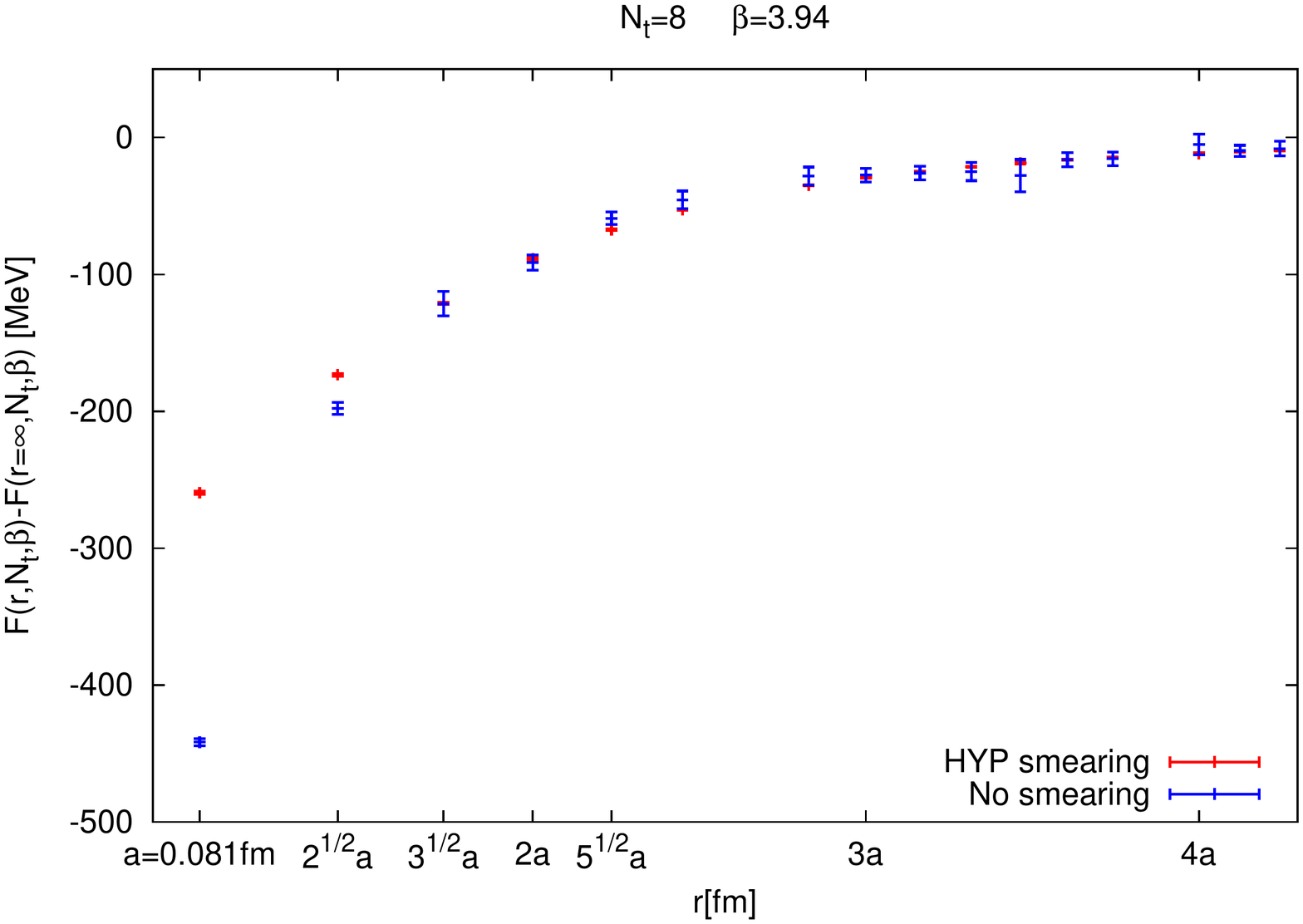}
\end{center}
\caption{The smeared and unsmeared free energies at a given $\beta$ and $N_t$, after the first step of the renormalization procedure. }
\label{fig:smear}
\end{figure}

Let us first mention that this Polyakov loop correlator behaves similarly to the baryon 
correlators in imaginary time do: at large values of $r$ we can get negative values of $C$ at some 
configurations\footnote{Of course, the ensemble average should in principle be positive definite.}. 
For this reason, it is highly desirable to use gauge field smearing which makes for a much better behavior at large $r$, at the
expense of unphysical behavior at small $r$. For this reason, we measured the correlators both without and with HYP smearing.
We expect that outside the smearing range (i.e. $r \geq 2a$) the two correlators coincide. This is supported by Figure \ref{fig:smear}.
Therefore we use the smeared correlators for $r \geq 2a$ and the unsmeared ones for $r<2a$.

\begin{figure}[h!]
\begin{center}
{ \includegraphics[width=0.48\textwidth, angle=0]{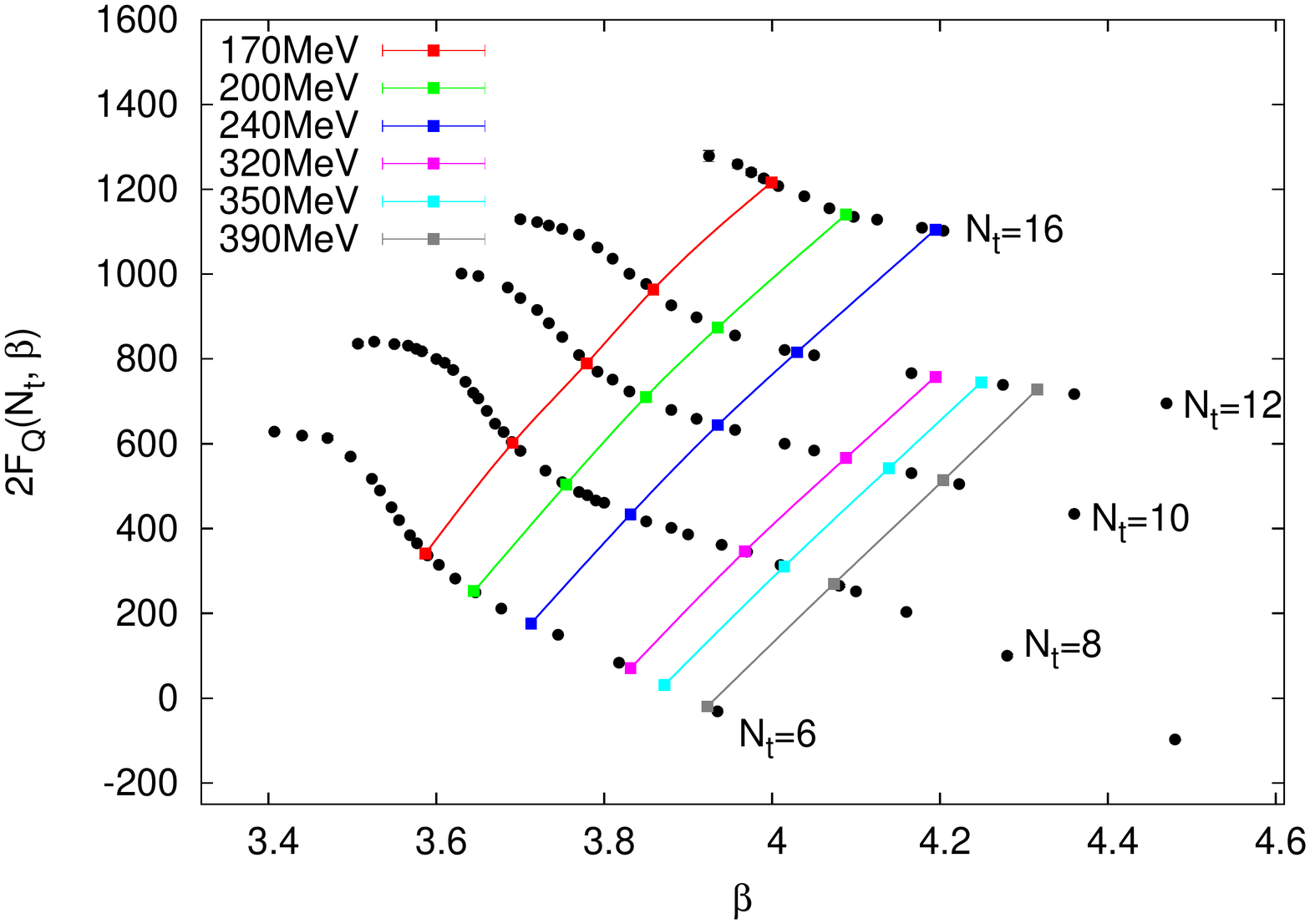} }
{ \includegraphics[width=0.48\textwidth, angle=0]{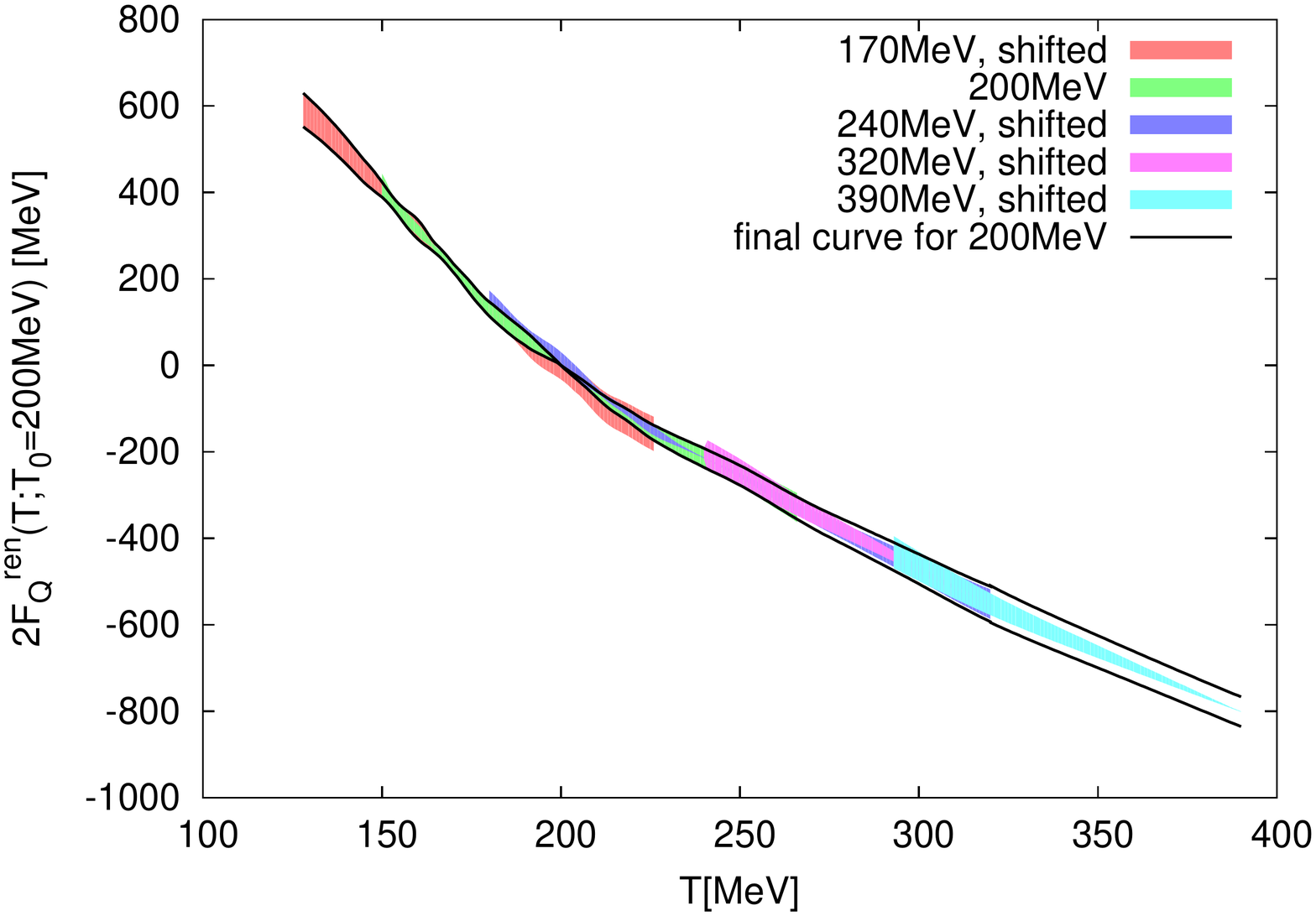} }
\end{center}
\caption{Left: Determining $F_Q (\beta,T_0)$ for different values of $T_0$ with interpolation. The bare $2F_Q$ values
for different values of $N_t$ are the black symbols. The colored symbols correspond to different fixed $T_0$ values for  each
$N_t$. The colored lines are interpolations between these points in $\beta$.
Right: $2 F_Q^{ren}(T;T_0)$ values in the continuum, calculated for
different values of $T_0$. For the final curve, all of the curves have been shifted to the position of the $T_0=200\rm{MeV}$ curve.
The errors of each piece decrease as we approach the corresponding $T_0$. For the final curve, we used linear error propagation,
assuming independent errors.
We also mention that calculating the continuum limit of $2 F_Q^{ren}(T;T_0)$ without HYP smearing leads to results consistent
with the one presented here.}
\label{fig:renorm_Nt_to_T0}
\end{figure}

\subsubsection{Single heavy quark free energy}

First, we discuss the implementation of the renormalization of the single heavy quark free energy, equation (\ref{eq:renorm1Q}).
Notice that if we implemented the renormalization condition (\ref{eq:renorm}) directly, then we would just need to
subtract $2 F_{Q}(\beta,T_0)$ from the unrenormalized free energy, so at first sight it looks like we are doing some unnecessary
rounds by doing this in two steps. What we gain by this is that we can extend the temperature range, at which we can do the continuum
limit. To understand this statement let us look at Figure~\ref{fig:renorm_Nt_to_T0}  (left). The dotted black symbols are bare values of $2F_Q$ at given
values of $N_t$ and $\beta$. The colored symbols are interpolations of these curves, in $\beta$ to the value of $\beta_0$ corresponding
to the temperature $T_0$ at each $N_t$. If we take for example $T_0=200\rm{MeV}$, corresponding to the green line in the figure, 
this gives us 5 points from the curve $F_Q(\beta,T_0)$. According to equation (\ref{eq:renorm1Q}) this is what we have to subtract
from the bare free energy at this value of $\beta$ to get the renormalized single quark free energy. The disadvantage of the green
curve, is that the $\beta$ range it covers is rather limited. So, if we want to be able to make a continuum limit from say the 
$N_t=8,10,12$ lattices, the temperature range we can cover is rather limited as well. The lowest temperature we will be able to
do a continuum limit at will be $(6/8) \times 200\rm{MeV} = 150\rm{MeV}$, and the highest temperature will 
be $(16/12) \times 200\rm{MeV} = 266\rm{MeV}$.To do a continuum limit at higher temperatures, we need 
the $F_Q(\beta,T_0)$ curve at higher values of $\beta$, and at first, it
looks as like that would need runs at higher values of $N_t$. This is not feasible, but there is a simple tricj to
extend the temperature range. Clearly, if we call the continuum limit of the single quark 
free energy
\begin{equation}
F_Q^{\rm{ren}}(T;T_0)= \lim_{\beta \to \infty}F_Q^{\rm{ren}}(\beta,T;T_0) \rm{,}
\end{equation}
than, for any value of $T$:
\begin{equation}
\label{eq:shift}
F_Q^{\rm{ren}}(T;T_0) - F_Q^{\rm{ren}}(T;T_1) = F_Q^{\rm{ren}}(T_1;T_0)
\end{equation}
is just a number\footnote{This statement is only true in the continuum. At finite lattice spacing there is also a lattice spacing 
dependent artifact in this difference.}. We can use this fact to extend the temperature range of the continuum limit by using different
values of $T_0$, that is different renormalization prescriptions, and shift them together by the value of the RHS of equ. (\ref{eq:shift}).
This is the procedure that we will follow. \\

To implement equation (\ref{eq:renorm1Q}), we first calculate $F_Q(\beta,N_t)$ or equivalently $F_{\bar{Q}Q} (r \to \infty,\beta,N_t)$ 
from equation (\ref{eq:F1Q}). Then at each $N_t$ we interpolate to the $\beta$ value corresponding to the temperature $T_0$, giving us some 
points of the  function $F_Q(\beta,T_0)$. Finally, we interpolate these $F_Q(\beta,T_0)$ points in $\beta$, obtaining 
the final curve we can use for the renormalization. This procedure is illustrated on Figure~\ref{fig:renorm_Nt_to_T0} (left). 
When doing this interpolation we take into 
account the error on the data points of $F_Q(\beta,N_t)$ 
via the jacknife method. The statistical errors of the single quark free energy are very
small, meaning that the interpolation method gives a comparable error to the final interpolated value. We estimate the systematic error of
the interpolations by constructing different interpolations. For interpolations of the $F_Q(\beta,N_t)$ curves we use linear 
and cubic spline interpolations (for each value of $N_t$), and for the interpolation of $F_Q(\beta,T_0)$ we use different polynomial 
interpolations(order 1,2), cubic spline and barycentric rational function interpolation. In total this means $2^5 \times 4=128$ different
interpolations, than for interpolating the bare $F_Q$ we use spline and linear interpolations, so for
the final renormalized values we have in total $128 \times 2 = 256$ different interpolations. All interpolations are 
taken to have the same weight. We use the median of this as the estimate, and the
symmetric median centered 68\% as the 1σ systematic error estimate~\cite{ScienceHadronSpectrum}. The statistical
and systematic errors turn out to be of the same order, and are than added in quadrature.
\\

After doing this procedure, the  $\beta$ range in which we can interpolate the $F_Q(\beta,T_0)$ curve is limited, therefore, the temperature 
range where we can do the continuum extrapolation is limited.  To extend the temperature range where we can calculate
the single heavy quark free energy, we use the fact that the single heavy quark free energies 
at different temperatures differ only by an additive constant in the  continuum. Therefore 
we use different values of $T_0$ to do the continuum extrapolation, and shift all those curves to the position of the 
$200\rm{MeV}$ curve. We used 5 different values of $T_0$, namely, $170$MeV, $200$MeV, $240$MeV, $320$MeV, and $390$MeV. 
The results of this analysis can be found in Figure \ref{fig:renorm_Nt_to_T0} (right). \\

For the continuum limits, we use the $N_t=8,10,12$ lattices, that are available at all temperatures. We use the $N_t=16$
lattice to estimate the systematic error of the continuum extrapolation, where it is available. If:
\begin{eqnarray*}
d_1=\left| \rm{cont.\;lim.}(8,10,12) \right| - \left| \rm{cont.\;lim.}(8,10,12,16) \right| \\
d_2=\left| \rm{cont.\;lim.}(8,10,12) \right| - \left| \rm{cont.\;lim.}(10,12,16) \right| \rm{,}
\end{eqnarray*}
then the systematic error of the continuum extrapolation is taken to be $\operatorname{max}\left(d_1,d_2\right)$. Where
the $N_t=16$ lattices are not available, 
we approximate  the relative systematic error by the average of the systematic errors at the parameter values where we had the
$N_t=16$ lattices available. This corresponds to an error level of approximately $10\%$. 
The systematic and statistical errors of the
continuum extrapolations are then added in quadrature. The linear 
fits of the continuum limit extrapolations all have good values of $\chi^2$. \\

Finally, we mention that the determination of the continuum limit of the Polyakov loop, or equivalently, that
single static quark free energy is already available in the literature. For two recent determinations of the Polyakov loop 
see Refs. ~\cite{Borsanyi:2010bp, Bazavov:2013yv}. The difference is that here we take the continuum limit at significantly 
higher temperatures.

\subsubsection{Heavy $\bar{Q}Q$ pair free energy}

Next, we turn to the determination of $\tilde{F}_{\bar{Q}Q}$ defined in equation~\ref{eq:Ftilde}. This quantity is UV finite, and goes 
to 0 as $r \to \infty$. Similarly to the single quark free energy, the determination of $\tilde{F}_{\bar{Q}Q}$ at a given value of $T$ 
and $r$ requires two interpolations. At first we are given $\tilde{F}_{\bar{Q}Q}$ at several values of $T$, at each $T$ we have a different 
value of the lattice spacing. If we want to know the value of $\tilde{F}_{\bar{Q}Q}$ at $(T,r)=(T^*,r^*)$ at some value of $N_t$, 
first we do an interpolation in the $r$ direction to the value $r^*$ at each given $T$, then we do an interpolation in the T direction, where the node
points for the interpolations are the interpolants in the previous step. The statistical error than can be estimated by constructing these
interpolations to every jacknife sample. For systematic error estimation we try different interpolations in the $r$ and $T$ directions.
In the r direction we have: polynomials of order 1,2,3,...,7 and a cubic spline, in the $T$ direction we have polynomials of order 1,2,3 and 
cubic spline. This is in total $4 \times 8 = 32$ different interpolations. Just as before, we use the median of these values as 
the estimate, and the symmetric median centered $68\%$ as the $1\sigma$ systematic error estimate. Like in the case of the 
single heavy quark free energies, the statistical and systematic errors turn out to be of the same order, and are then added in quadrature. \\

Next, we do the continuum extrapolation. Here we also take a similar approach as in the previous subsection.
For the continuum extrapolations, we use the $N_t=8,10,12$ lattices, that are available at all temperatures. We use the $N_t=16$
lattice to estimate the systematic error of the continuum extrapolation, exactly like before. Also, where the $N_t=16$
lattices are not available, we estimate the systematic error, as in the previous section, by the average of the systematic
error at the points where we do have $N_t=16$ lattices (approximately $7\%$). The linear fits of the continuum limit extrapolations all 
have good values of $\chi^2$. \\

\begin{figure}[t!]
\begin{center}
\includegraphics[width=0.7\textwidth, angle=0]{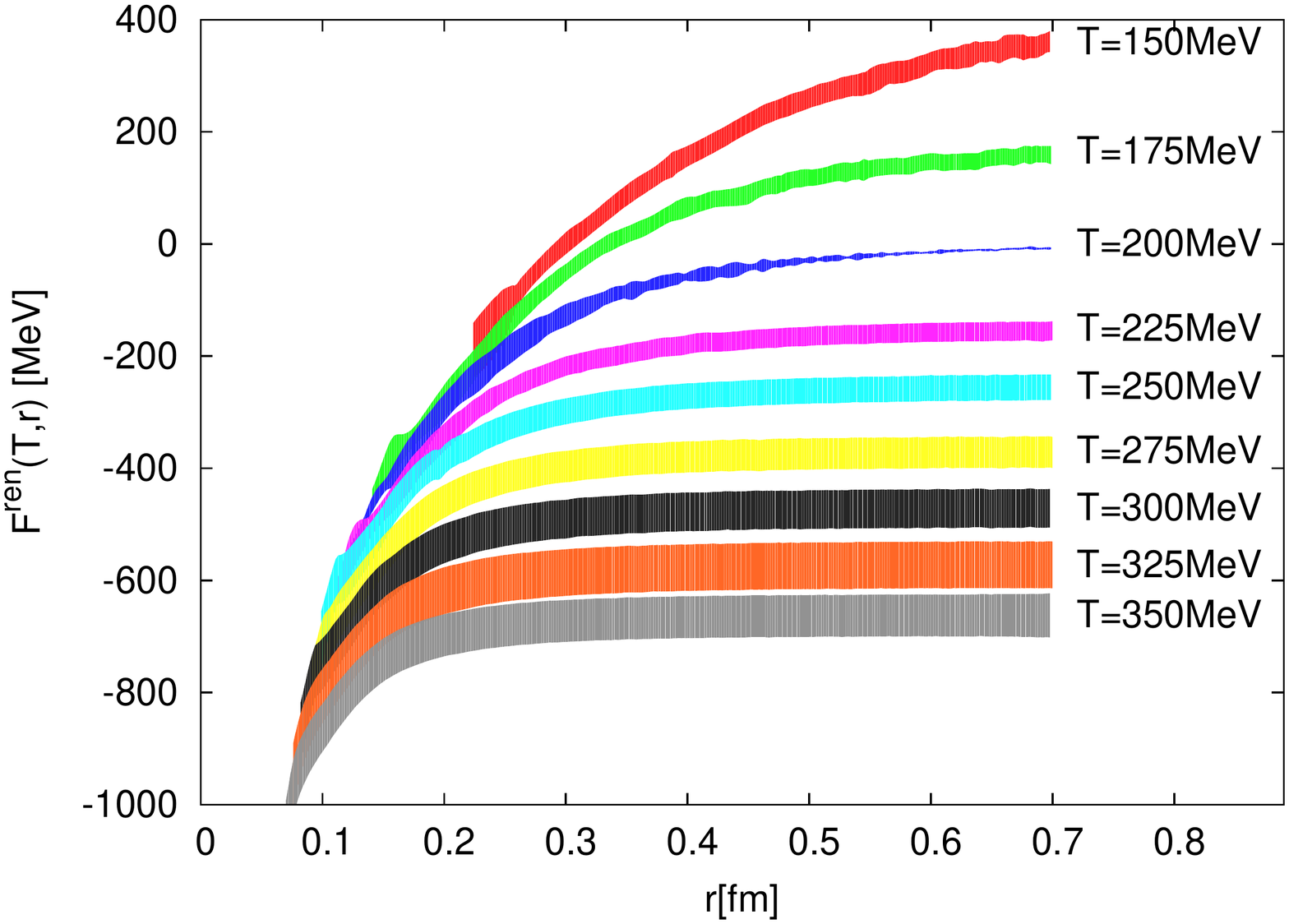}
\caption{Continuum values of the static $\bar{Q}Q$ free energy at different temperatures. Note that the curves seem to tend to
the same curve as $r \to 0$, corresponding to the expectation that UV physics is temperature independent. }
\label{fig:final_free_energy}
\end{center}
\end{figure}

Next, we add the values of $2F_Q$, determined in the previous subsection, and visible in Figure \ref{fig:renorm_Nt_to_T0} 
to the free energy values to obtain the final results in Figure \ref{fig:final_free_energy} (errors are added in quadrature). 
Note, that the $N_t=6$ lattices were only used in the whole analysis to extend the $\beta$ range of the renormalization 
condition for the single quark free energy.

\section{Magnetic and electric screening masses}
\begin{figure}
\begin{center}
\includegraphics[angle=0, width=0.7\linewidth]{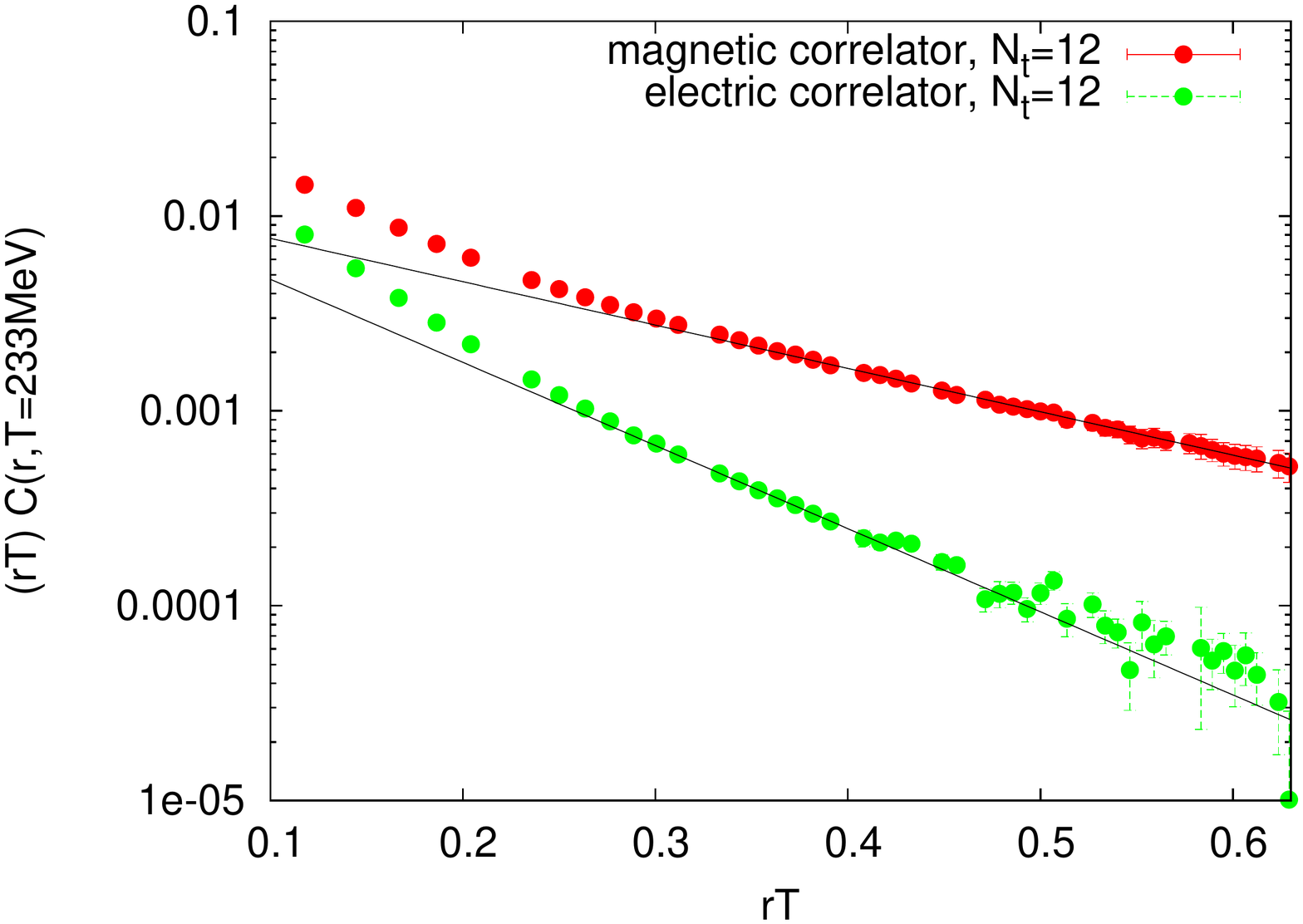}
\end{center}
\caption{Illustrating the dominance of the magnetic contribution (or $m_E>m_M$) with the lattice data at $N_t=12$ ($32^3 \time 12$ lattices with
HYP smearing). It can be clearly seen 
that the electric correlator drops faster. To lead the eye, we included on the plot the Yukawa fits to the tails of the correlators.}
\label{fig:mEvsmM}
\end{figure}

\begin{figure}
\includegraphics[angle=0, width=0.48\linewidth]{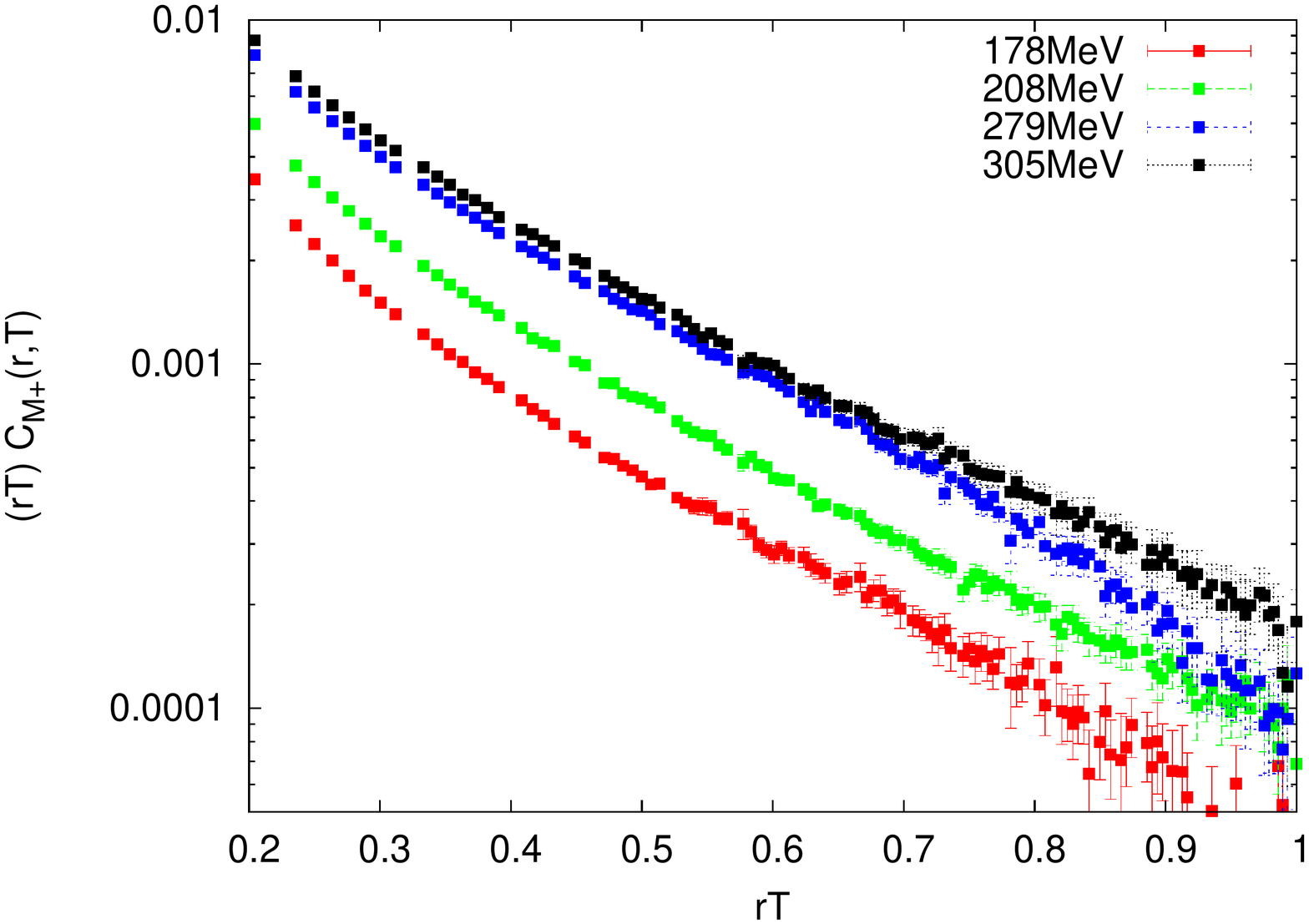}
\includegraphics[angle=0, width=0.48\linewidth]{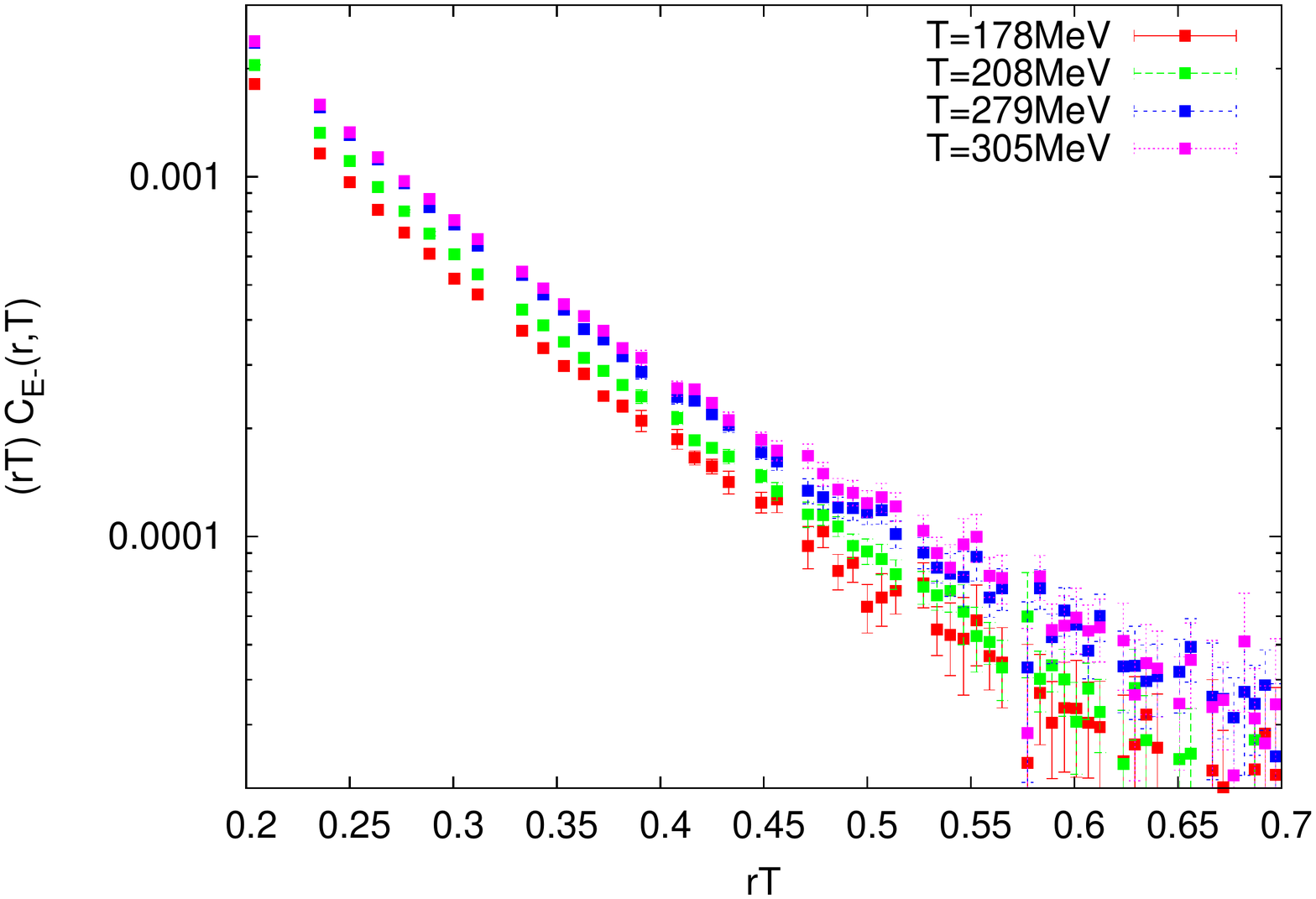}
\caption{Illustrating that the screening masses are approximately proportional 
to the temperature. Since the x axis of this plot is $rT$, if one 
assumes a Yukawa form of the correlator, than 
the slopes of these curves are just $m_M/T$ and $m_E/T$ respectively. The 
fact that the graphs are approximately parallel straight lines
suggests that these ansatzes are approximately correct, and that the masses are approximately
proportional to the temperature.}
\label{fig:mproptoT}
\end{figure}

We continue with the discussion of the electric and magnetic screening masses obtained from the correlators 
(\ref{eq:Cmag}) and (\ref{eq:Cel}). For this analysis we only use lattices above the (pseudo)critical
temperature, since that is the physically interesting range for screening. Next, we mention that for this analysis,
we only use the data with HYP smearing, since we are especially interested in the large r behavior. Before 
going on to the actual fitting procedure of the screening masses let us first illustrate some simple relations, 
with the raw lattice data of the electric and magnetic correlators. First $C_{E-}(r,T) \ll C_{M+}(r,T)$ as $r \to \infty$, 
or equivalently, that the electric screening mass is larger than the magnetic one. This can be seen on Figure \ref{fig:mEvsmM}. 
The next observation is that the screening masses in both channels are approximately proportional to the temperature. 
This can be seen on Figure \ref{fig:mproptoT}. Both of these facts are expected to hold at high temperatures, but 
these lattice results suggest that they hold at lower
temperatures as well.\\

Next, we turn to the actual determination of the screening masses. 
So far there has been one determination of electric and magnetic screening masses on the lattice 
using the non-perturbative definition given by ref. ~\cite{Arnold:1995bh}. That study used 
2 flavours of Wilson fermions with a somewhat heavy pion, and did not attempt a continuum extrapolation~\cite{Maezawa:2010vj}. \\

\begin{table}[t!]
\centering
\begin{tabular}{ | c | c | c | c |}
  \hline
  Correlator type & $(rT)_{\rm{min}}$ & $(rT)_{\rm{max}}$ & $Pr \left( \rm{KS},\rm{uniform} \right)$ \\
  \hline
  Magnetic        & $0.43$              & $0.9$               & 0.007 \\ 
  Magnetic        & $0.45$              & $0.9$               & 0.016 \\ 
  Magnetic        & $0.465$             & $0.9$               & 0.30  \\ 
  Magnetic        & $0.5$               & $0.9$               & 0.38  \\ 
  Magnetic        & $0.61$              & $0.9$               & 0.96  \\ 
  \hline
  Electric        & $0.3$               & $0.65$              & $3 \cdot 10^{-7}$ \\
  Electric        & $0.32$              & $0.65$              & 0.018 \\ 
  Electric        & $0.35$              & $0.65$              & 0.31  \\ 
  Electric        & $0.43$              & $0.65$              & 0.94  \\ 
  \hline
\end{tabular}

\caption{Hypothesis testing, using fits at all values of $N_t=8,10,12$ and all values of $\beta$. This means 33 sampled 
values in total, with fixed values of the low range of the fit $(rT)_{\rm{min}}$. 
One can see a rather sharp increase
in the probabilities for the magnetic correlator at $(rT)_{\rm{min}}=0.465$ and for the electric correlator at $(rT)_{\rm{min}}=0.35$.
This table justifies our choice for the ranges of $(rT)_{\rm{min}}$ values used in our systematic error estimation. }
\label{table:KS}
\end{table}

Since the masses are expected to be proportional to the temperature, the natural distance unit in 
this problem is $rT$, so we give limits on the range of the fits in these units. For the correct
determination of the screening masses, special care is needed in the choice of the fit interval.
To find the proper minimum $rT$ value of the fits, we use hypothesis testing, similar to
that in Ref.~\cite{Borsanyi:2014jba}. If the fits 
are good, than the value of $\chi^2$, defined as:
\begin{equation}
\chi^2 = \sum_{i,j} (C^{\rm{fit}}_i - C^{\rm{data}}_i) \bm{\mathcal{C}}_{ij}^{-1} (C^{\rm{fit}}_j - C^{\rm{data}}_j) \rm{,}
\end{equation}
should have a $\chi^2$ distribution, with the appropriate degrees of freedom. Here $\bm{\mathcal{C}}_{ij}$ is the covariance matrix. 
In this case the quantity
\begin{equation}
Q = \int_{\chi^2}^{\infty} \left( \rm{Probability \; density \; of \; \chi^2} \right)(x) dx \rm{,}
\end{equation}
should have a uniform distribution on $[0,1]$. If we fix the range of all the fits in $rT$ units, each fit
(at some value of $N_t$ and $\beta$) gives one pick from a supposed uniform distribution in Q. This is
equivalent to having multiple picks from the same uniform distribution. We will test this hypothesis with
 a Kolmogorov-Smirnov test for the uniform distribution. Here one determines the maximum value of the absolute 
 difference between the expected and measured cumulative probability distributions. This is then used to define a significance 
 level or probability that the measured distribution can indeed be one originating from the expected uniform distribution.
 These probabilities are listed in Table \ref{table:KS}. We will only use value of $(rT)_{\rm{min}}$ where the Kolmogorov
 probability is at least $0.3$. This test tells us, that for systematic error estimation, we will have, for the magnetic 
correlator $(rT)_{\rm{min}}$ going from 0.465 to 0.61,
and for the electric correlator we have $(rT)_{\rm{min}}$ going from 0.35 to 0.43
\footnote{$(rT)_{\rm{max}}$ was fixed on both cases. Increasing $(rT)_{\rm{max}}$ results in
a less precise covariance matrix and correspondingly, somewhat worse $\chi^2$ values, but consistent screening masses.
For example, if for the magnetic correlator we choose $(rT)_{\rm{max}}=1$ instead of $0.9$, the final value of the 
Kolmogorov-Smirnov probability in Table \ref{table:KS} will not be $96\%$, but $38\%$ instead. Nevertheless the growing trend 
in the probabilities will be the same. Also, we will get the same results within uncertainties.}.\\

At this point we mention that for the continuum limit we will not use the $N_t=16$ lattices, because the
mass fits there have huge error bars.  Nevertheless, when the continuum limit is done, we will see that
the values of the masses at the $N_t=16$ lattices are consistent with the continuum estimates. Also, if we use them, 
we get the same results, because they do not give a contribution to the continuum limit, due to the big errors.\\

\begin{figure}[t!]
\includegraphics[angle=0, width=0.48\linewidth]{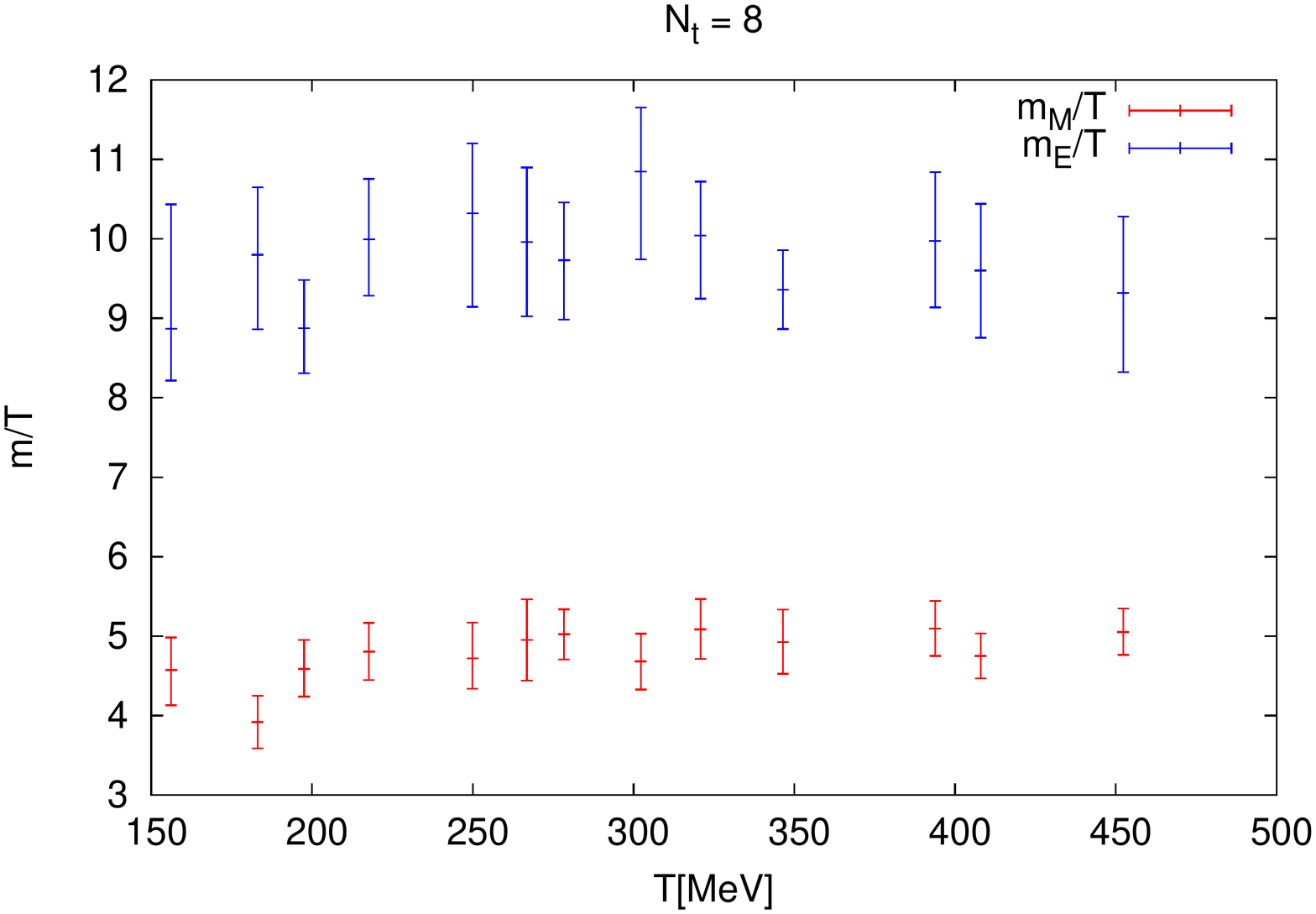}
\includegraphics[angle=0, width=0.48\linewidth]{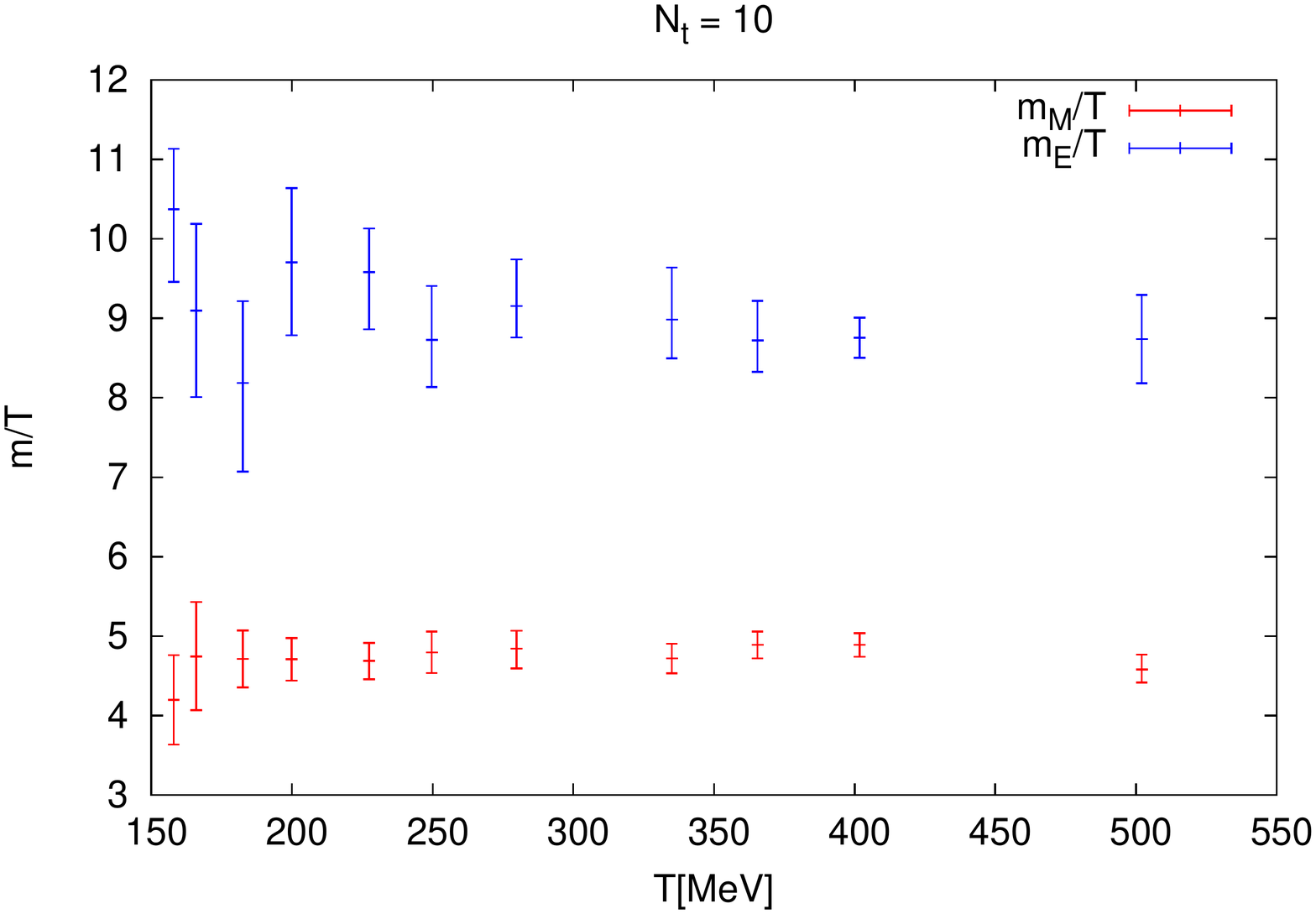} \\
\includegraphics[angle=0, width=0.48\linewidth]{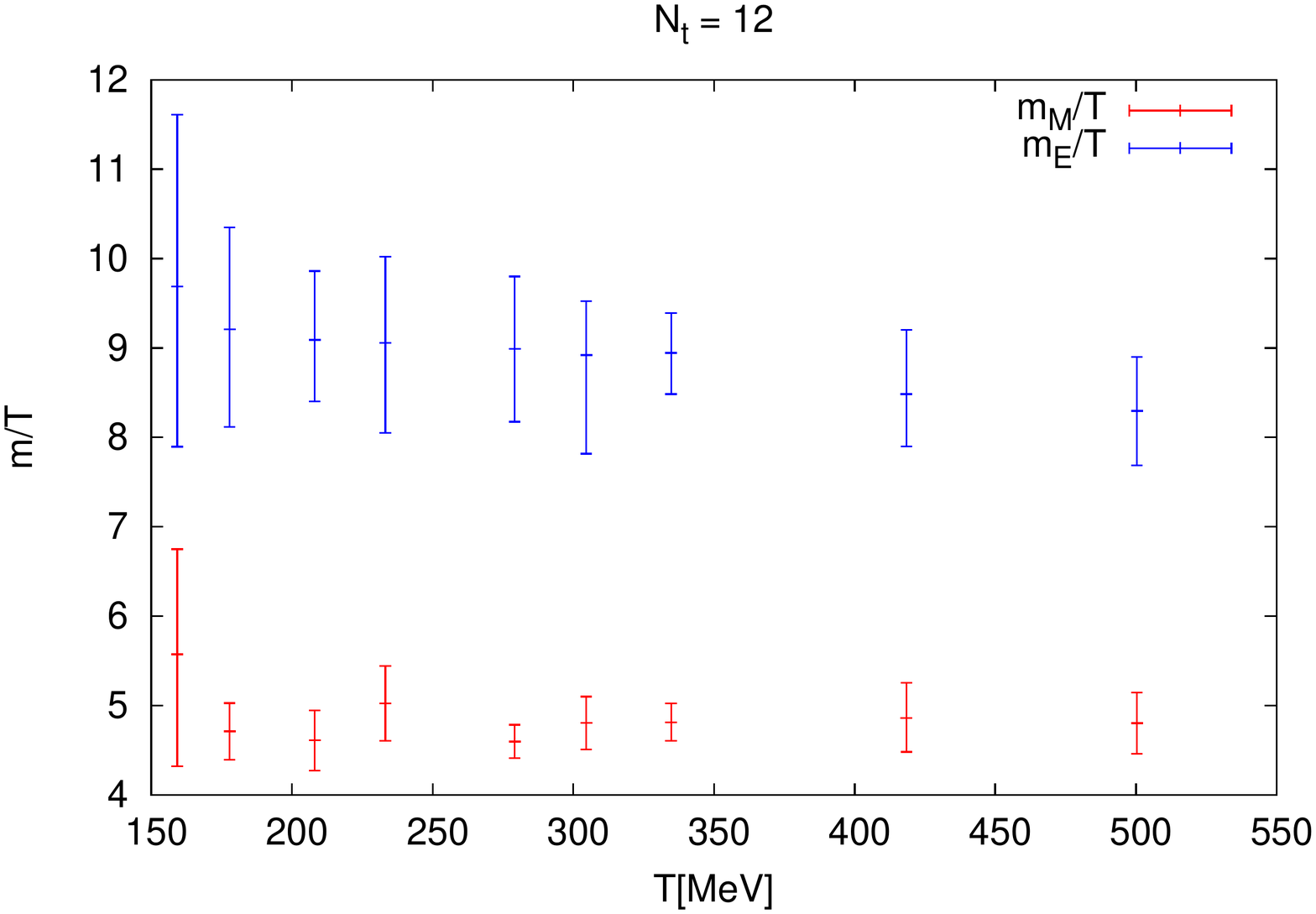}
\includegraphics[angle=0, width=0.48\linewidth]{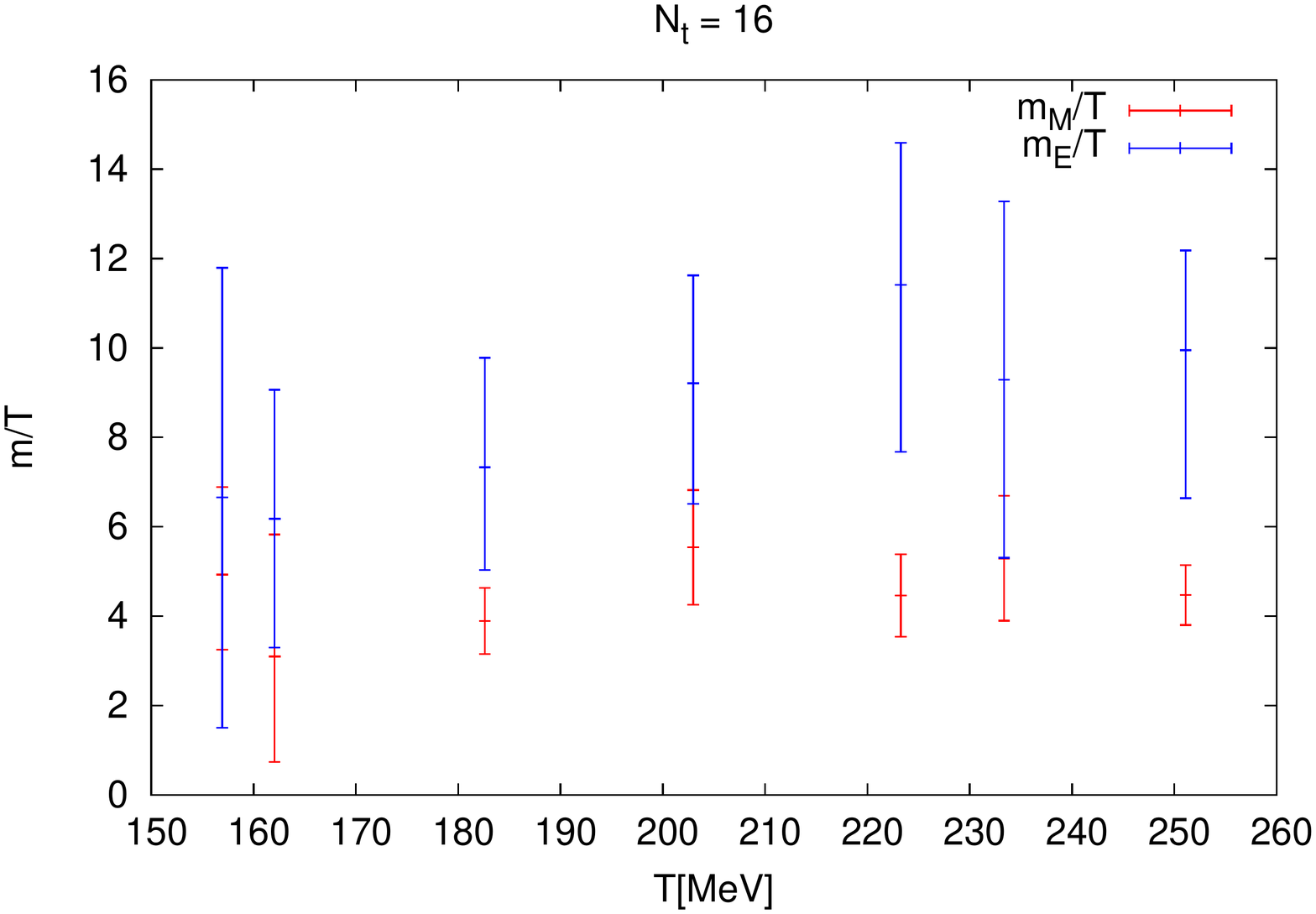}
\caption{The fitted values of electric and magnetic screening masses at the different values of $N_t$. 
} 
\label{fig:mass_fits}
\end{figure}

Now that we have estimated the proper $rT$ range of the fits, we go on to the fitting of the masses. 
The results of the fits at different
values of $N_t$ can be seen in Figure \ref{fig:mass_fits}.
The systematic errors come from changing the lower limit of the fit, in the case of the magnetic 
correlator, from $(rT)_{\rm{min}}=0.465$ to $(rT)_{\rm{min}}=0.61$, and in the case of the electric correlator, from $(rT)_{\rm{min}}=0.35$ to
$(rT)_{\rm{min}}=0.43$. The results coming from different values of $(rT)_{\rm{min}}$ are weighted using
the Akaike Information Criterion(AIC) \cite{Akaike:1974}. The median of the weighted histogram gives the central value,
and the central $68\%$ the systematic error estimate. Note that using the Q values as weights or uniform weights 
gives a very similar result.
The statistical error comes from a jacknife analysis with 20 jacknife samples. 
The two errors turn out to be of similar magnitude (with the statistical error being somewhat bigger) and are then added 
in quadrature. \\

Next, we fit linear functions to all screening masses at all values of $N_t$,
and use these to do a continuum extrapolation from the $N_t=8,10,12$ lattices.
Taking into account the errors of the linear fits,  all $\chi^2$ values of the continuum limits are very good. The continuum limit, in addition to the statistical
error, also has a systematic error estimated, from doing a 2 point linear extrapolation from the $N_t=12,10$ lattices, and taking the
difference of the extrapolated value from fitted value to the $N_t=8,10,12$ lattices\footnote{In the previous section, we used 
the $N_t=16$ lattices for systematic error estimation, here however, we do not use them since they do not improve the
statistical accuracy of the continuum limits.}. The statistical and systematic errors are added in quadrature. \\

\begin{figure}
\includegraphics[angle=0, width=0.48\linewidth]{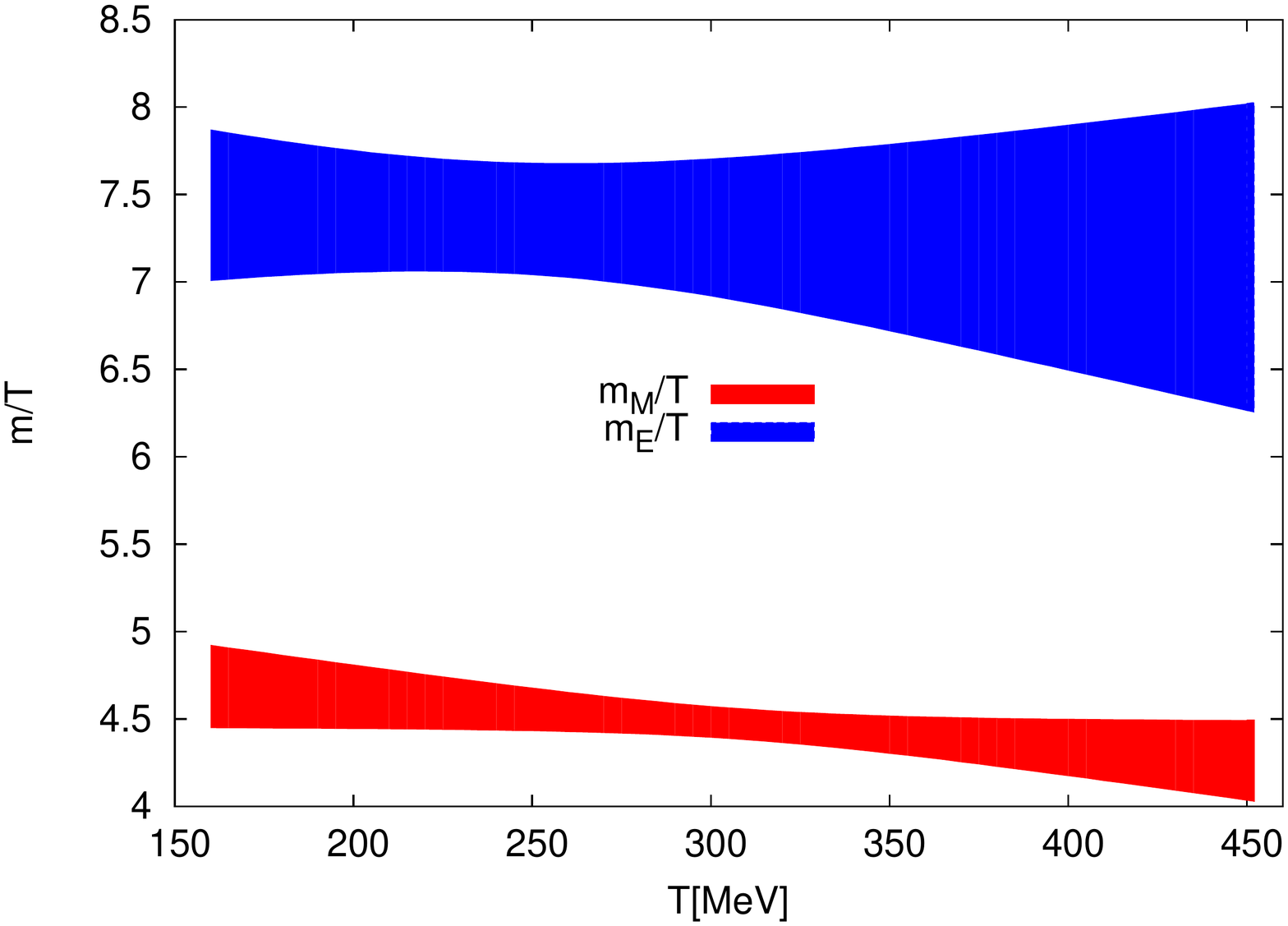}
\includegraphics[angle=0, width=0.48\linewidth]{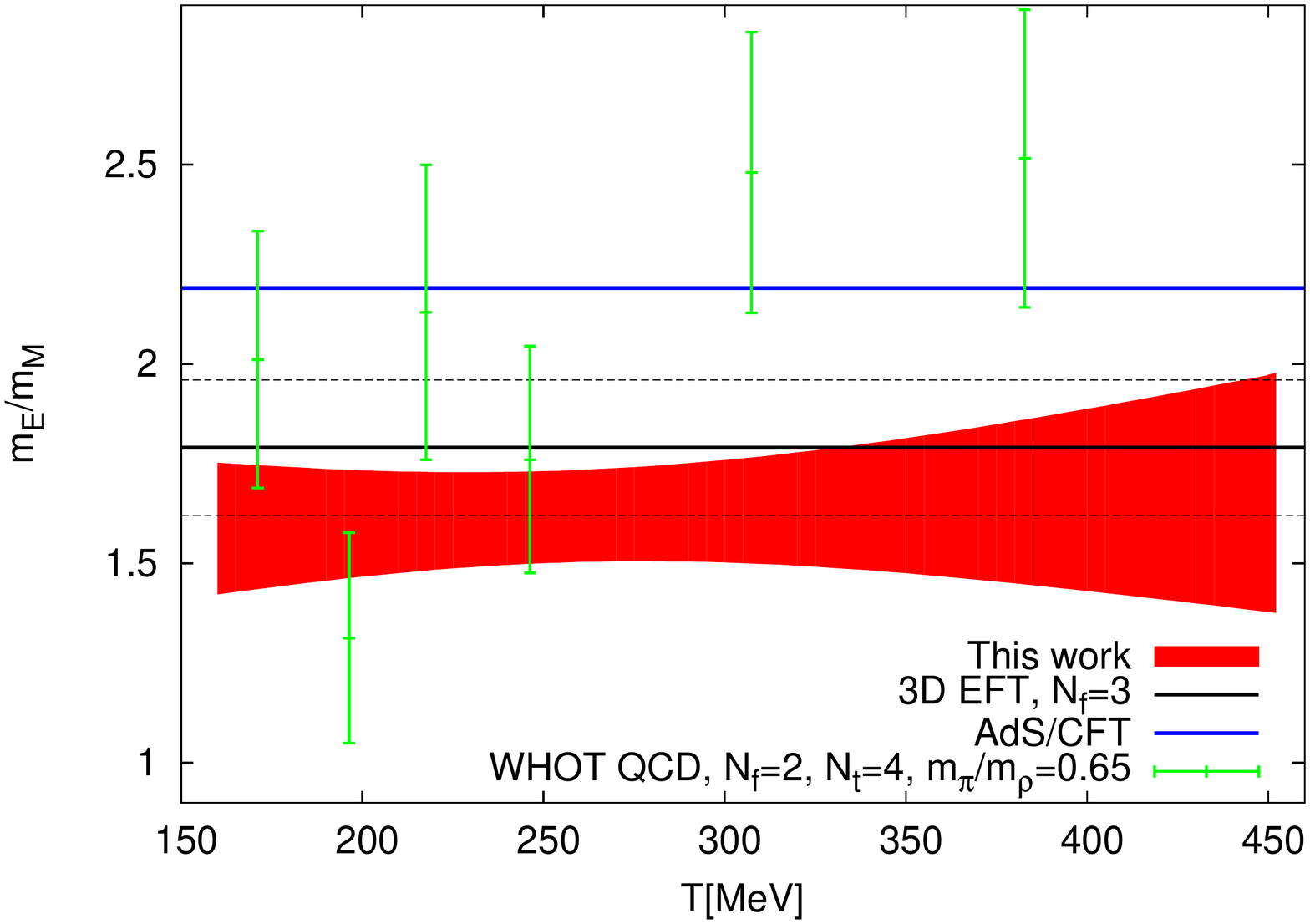}
\caption{The continuum extrapolations of the screening masses and the ratio of the screening masses.
For the ratio $m_E/m_M$ we also included different estimates from the literature:
Lattice results from Ref.~\cite{Maezawa:2010vj}, dimensionally reduced 3D effective
field theory results from Ref.~\cite{Hart:2000ha}, and results from
$\mathcal{N}=4$ SYM plasma with AdS/CFT from Ref.~\cite{Bak:2007fk}.
}
\label{fig:mass_contlim}
\end{figure}

We finish this section by comparing our results to those from earlier approximations in the 
literature. For comparison let us use our results at $T=300\rm{MeV} \approx 2 T_c$. Here we have:
\begin{itemize} 

\item This work: 2+1 flavour lattice QCD at the physical point after continuum extrapolation:
\begin{eqnarray*}
m_E/T=7.31(25)    \quad m_M/T=4.48(9)  \\
m_E/m_M=1.63(8)
\end{eqnarray*}

\item Ref.~\cite{Maezawa:2010vj}: 2 flavour lattice QCD with Wilson quarks, 
a somewhat heavy pion $m_{\pi}/m_{\rho}=0.65$, no continuum extrapolation
\begin{eqnarray*}
m_E/T=13.0(11)    \quad m_M/T=5.8(2)  \\
m_E/m_M=2.3(3)
\end{eqnarray*}

\item From Table 1 of Ref.~\cite{Bak:2007fk}: $\mathcal{N}=4$ SYM, large $N_c$ limit, AdS/CFT 
\begin{eqnarray*}
m_E/T=16.05    \quad m_M/T=7.34  \\
m_E/m_M=2.19
\end{eqnarray*}

\item From Figure 3 of Ref.~\cite{Hart:2000ha}: dimensionally reduced 3D effective theory, $N_f=2$ massless quarks
\begin{eqnarray*}
m_E/T=7.0(3)   \quad m_M/T=3.9(2)  \\
m_E/m_M=1.79(17)
\end{eqnarray*}

\item From Figure 3 of Ref.~\cite{Hart:2000ha}: dimensionally reduced 3D effective theory, $N_f=3$ massless quarks
\begin{eqnarray*}
m_E/T=7.9(4)    \quad m_M/T=4.5(2)  \\
m_E/m_M=1.76(17)
\end{eqnarray*}

\end{itemize}

We note, that our results are closest to the results from dimensionally reduced effective field theory.

\section{Conclusions}

In this paper we have determined the renormalized static quark-antiquark free energies 
in the continuum limit. We introduced a two step renormalization procedure using only the finite 
temperature results. The low radius part of the free energies tended to the same curve,
corresponding to the expectation that at small distances, the physics is temperature 
independent. We also calculated the magnetic and electric screening masses, from the real 
and imaginary parts of the Polyakov loop respectively. As expected, both of
these masses approximately scale with the temperature as $m \propto T$, with $m_M<m_E$, therefore, magnetic 
contributions dominating at high distances. The values we got for the screening masses are close to 
the values from dimensionally reduced effective field theory.

\section*{Acknowledgment}
Computations were carried out on GPU clusters \cite{Egri:2006zm} at the
Universities of Wuppertal and Budapest as well as on supercomputers in
Forschungszentrum Juelich. \\

This work was supported by the DFG Grant SFB/TRR 55, ERC no. 208740. and 
the Lendulet program of HAS (LP2012-44/2012).\\

\bibliography{cikk_T}{}

\end{document}